\documentclass[twocolumn, numberedappendix, superscriptaddress, nofootinbib]{revtex4-2}

\usepackage{graphicx}
\usepackage{dcolumn}
\usepackage{amssymb, amsmath}
\usepackage{siunitx}
\usepackage{mathtools}

\usepackage[T1]{fontenc}
\usepackage{newpxtext}
\usepackage{euler}

\usepackage[colorlinks=true, 
            linkcolor=blue, 
            urlcolor=blue, 
            citecolor=blue]{hyperref}

\usepackage{orcidlink}

\usepackage{xcolor}
\usepackage{xspace}
\def\Sref#1{Sec.~\ref{#1}\xspace}
\def\Fref#1{Fig.~\ref{#1}\xspace}
\def\Eref#1{Eq.~\eqref{#1}\xspace}

\newcommand{\mcm}{\mathcal{M}}
\newcommand{\mcp}{\mathcal{P}}
\newcommand{\mcr}{\mathcal{R}}

\begin{document}

\title{How much can shape degeneracies affect the Hubble constant measurement?}

\author{Ashish K. Meena\orcidlink{0000-0002-7876-4321}}
\email{akm@iisc.ac.in}
\affiliation{Department of Physics, Indian Institute of Science, Bengaluru 560012, India}

\author{Prasenjit Saha\orcidlink{0000-0003-0136-2153}}
\email{psaha@physik.uzh.ch}
\affiliation{Physik-Institut, University of Zurich, Winterthurerstrasse 190, CH-8057 Zurich, Switzerland}

\author{Liliya L. R. Williams\orcidlink{0000-0002-6039-8706}}
\affiliation{Minnesota Institute for Astrophysics, University of Minnesota, 116 Church Street SE, Minneapolis, MN 55455, USA}

\author{Derek Perera\orcidlink{0000-0002-4693-0700}}
\affiliation{Minnesota Institute for Astrophysics, University of Minnesota, 116 Church Street SE, Minneapolis, MN 55455, USA}

\begin{abstract}
Cluster lens models are constrained almost entirely by observed 
lensed image positions, leaving a large space of mass perturbations
with arbitrary amplitudes such that they reproduce every image 
exactly. 
Such degeneracies are known as shape degeneracies~(ShaDes). We show 
that these ShaDes shift the observed time delays, by the 
difference of the perturbing potential between the two images, 
without depending on the number of source/lens planes.
Bounding their amplitude by preserving the positivity of the 
convergence, image multiplicity, and parity, in MACS\,J0138.0$-$2155, 
they can contribute at least $\simeq7$~days to the delay between the 
detected images of SN-Encore (i.e., $17.5\%$ of the measured value 
of~$39.8$~days) and $\simeq112$~days (i.e., $2.8$--$3.7\%$ of 
predicted value of~$3,000-4000$~days) to the 
reappearance intervals predicted for SN-Encore and SN-Requiem. These 
error bars transfer one-to-one to $H_0$, and remain invisible in 
comparisons of independent models.
\end{abstract}

\maketitle

\section{Introduction}
\label{sec:intro}
The discovery of multiply imaged supernovae~(SNe) behind clusters --- 
SN-Refsdal~\citep{2016ApJ...819L...8K, 2023Sci...380.1322K}, 
SN-Requiem~\citep{2021NatAs...5.1118R}, 
SN-H0pe~\citep{2024ApJ...961..171F, 2025ApJ...979...13P}, and 
SN-Encore~\citep{2026ApJ...998..219P, 2026A&A...708A.291S} --- 
has turned cluster lensing into a standard route to the Hubble 
constant measurement.
Because the inference is geometric, it is independent of both the 
distance ladder and the CMB. 
The measurement is conceptually simple: a supernova (or any 
time-varying source) appears in multiple lensed images of the same 
source, the relative arrival times are measured, and a mass model
converts those times into a distance, hence into $H_0$. 
For example, in the case of SN-Encore (and SN-Requiem), detected in 
MACS\,J0138.0$-$2155, seven teams used six different 
lens modeling packages (covering parametric, free-form, and hybrid
methods) to model the galaxy cluster, along with a 
strict blind protocol in which no results were exchanged between 
teams until every model was finalised~\citep{2026A&A...708A.291S}. 
The above exercise is the current state of the art.
Many more strongly lensed SNe in cluster lenses are expected in 
the near future, both from new discoveries and from the reappearance
of images of SNe already seen, e.g.,~\cite{2026arXiv260930440F}.
 
The constraints available to a cluster lens model are primarily the 
multiply lensed image positions. Magnification ratios (or flux ratios) 
are polluted by microlensing, millilensing, differential extinction, 
and intrinsic source structure and are rarely used as constraints. 
Stellar kinematics constrain individual members, not the global mass 
sheet. Weak lensing~\cite{2001PhR...340..291B} and 
X-ray data~\cite{2013SSRv..177..119E} constrain mass at resolutions
far coarser than the separation between individual lensed images. 
In practice, therefore, the likelihood used by every strong lens 
modeling team depends on the observed image positions alone. 

The current work is about systematics that remain invisible to such
position-based lens models, namely, the \emph{shape 
degeneracies}~\citep{2006ApJ...653..936S, Williams2026}. 
Shape degeneracies can be constructed such that at any arbitrary 
amplitude, they leave the observed image positions unchanged, although 
they do change the observed time delay between multiple images. 
Below, following~\cite{2006ApJ...653..936S, Williams2026}, we define
this degeneracy class and study its implications for SN-Encore and
SN-Requiem in MACS\,J0138.0$-$2155~\cite{2026A&A...708A.291S}.

\section{Shape degeneracies}
\label{sec:shades}

\subsection{Definition}
Let us assume that we have a best-fit strong lens model for a given
galaxy cluster and denote it by~$\mcm$. Let $\mcp$ be a perturbation 
added to this best-fit model $\mcm$, with deflection 
$\pmb{\alpha}_\mcp$, such that,
\begin{equation}
    \pmb{\alpha}_\mcp(\pmb{\theta}_i) = \mathbf{0}, \qquad i = 1,\dots,N,
    \label{eq:defining}
\end{equation}
where~$\pmb{\theta}_i$ is the multiply lensed image positions. 
As a result, the perturbed model $\mcm+\mcp$ maps every image to 
\emph{exactly} the same source position as $\mcm$ does, leaving the
corresponding likelihood (which is maximized during lens modeling) 
unchanged. Since \Eref{eq:defining} is 
linear and homogeneous in $\mcp$, any multiple $\lambda \mcp$ 
satisfies it too. Hence, the data constrain neither the presence nor 
the amplitude of $\mcp$. Such perturbations are known as shape
degeneracies~(ShaDes)\footnote{The coincidence with the Shades of 
Ankh-Morpork is entirely apt: a notorious district, unmapped, 
unpoliced, and rather busier than the authorities suppose.}. 
We note that the above definition is a restrictive
version of the definition provided in \citet{Williams2026}, where 
one can have constant deflection at the image positions and move the
source accordingly. Our definition provides a conservative estimate 
of errors in time delay measurements.
The effect of constant deflection is further discussed in 
\Sref{ssec:constant}.
From the above definition, we can see that
to construct ShaDes, we do not need a strong lens mass model a 
priori, and we can simply use the observed image positions. However, 
as we will see below, a strong lens model is required to set 
physically motivated constraints on their amplitudes.

ShaDes are distinct from the well-known mass-sheet 
degeneracy \citep[MSD;][]{1985ApJ...289L...1F, 1988ApJ...327..693G, 
2000AJ....120.1654S} 
and the source-position 
transformation~\cite[SPT;][]{2014A&A...564A.103S}.
The MSD and SPT are \emph{global} transformations, i.e., they rescale 
the source plane and are therefore, in principle, breakable with 
measurements such as the velocity dispersion. ShaDes are different 
in nature as they are \emph{local and configuration-specific}. 
It is also important to note that MSD is not a subset of the 
\emph{above-defined} ShaDes. 
In MSD, the deflection angles at the observed image positions are 
modified, which is not the case here.
On the other hand, the monopole 
degeneracy~\cite{2000AJ....120.1654S, 2008MNRAS.389..415L, 
2012MNRAS.425.1772L} is a proper subset.

\subsection{Construction}
\label{ssec:shades_construct}
To construct the ShaDes, we adopt the following procedure. 
We start by expanding the perturbing 
potential~$\psi_\mcp = \Sigma_{j=1}^{N_c} m_j \psi_j$ in a basis of 
$N_c$ compact components. With that, \Eref{eq:defining} becomes,
\begin{equation}
    \mathsf{A}\,\pmb{m}=\mathbf{0},\qquad
    \mathsf{A}\in\mathbb{R}^{2N\times N_c},
    \label{eq:null}
\end{equation}
with column $j$ of $\mathsf{A}$ holding the deflection of a unit-mass 
component at all images. The degeneracy is $\ker[\mathsf{A}]$, 
obtained from the singular value decomposition~(SVD) 
$\mathsf{A}=\mathsf{U}\mathsf{S}\mathsf{V}^{\!\top}$, of dimension
$n_{\rm free}=N_c-\mathrm{rank}[\mathsf{A}]$. To construct the
degeneracy, we use Plummer components~\cite{1911MNRAS..71..460P},
\begin{equation}
    \psi_j(\pmb{\theta}) = \frac{2\rm{G}}{c^2} \frac{1}{D_d} 
    \ln\left(|\pmb{\theta}-\pmb{\theta}_{c,j}|^2+\theta_{s,j}^2\right),
\label{eq:plummer}
\end{equation}
with $\pmb{\theta}_{c,j}$ and $\theta_{s,j}$ representing 
the center and angular core radius for $j$-th component,
laid on a lattice of spacing $s$ inside the smallest ellipse
enclosing the images, with cores $\theta_{s,j}=1.5\,s$, where
$s$ represents the separation between basis components.
$D_d$ represents the angular diameter distance to the lens. 
Writing $a,b$ for the ellipse semi-axes, the lattice holds 
$N_c\simeq\pi ab/s^2$ components, so that
\begin{equation}
n_{\rm free}\gtrsim 2N\left[(s_{\rm crit}/s)^2-1\right],
\qquad s_{\rm crit}\equiv\sqrt{\pi ab/2N},
\label{eq:dim}
\end{equation}
and the degeneracy opens up once the basis is finer than 
$s_{\rm crit}$. For our current analysis, we fix $s=s_{\rm crit}/2$.
The effect of variation in~$s$ is further discussed in
\Sref{sec:shades_basis}.

\subsection{Amplitude bound} 
\label{ssec:shades_bound}
Since the data does not constrain the amplitude ($\lambda$) at all, 
the bound must be physically motivated. In our current work, we impose 
three requirements. 

First, the total convergence of the perturbed model must
remain non-negative, 
$\kappa_{\mcm+\mcp} = \kappa_\mcm + \lambda\,\Delta\kappa \geq 0$, 
wherever the best-fit model has mass. If $\Delta\kappa \geq 0$, the
above condition holds for any $\lambda$. On the other hand, 
if $\Delta\kappa < 0$, it requires 
$\lambda \leq \kappa_\mcm/(-\Delta\kappa)$, so the largest admissible 
amplitude is
\begin{equation}
    f_+=\min_{\pmb{\theta}:\,\kappa_\mcm>0,\ \Delta\kappa<0}\ 
    \frac{\kappa_\mcm(\pmb{\theta})}{-\Delta\kappa(\pmb{\theta})}.
    \label{eq:pos}
\end{equation}
While applying the $\kappa_\mcm>0$ restriction, we excludes pixels 
at which the best-fit model itself returns a vanishing or negative 
convergence. Such pixels can only arise from the parametrization 
(such as sharp truncation in the density profile) rather than from 
the data, at least for the parametric models. 
The bound is not sensitive to the extent of the image plane grid.
The basis components are centered within the ellipse enclosing the 
images and their convergence fall as $r^{-4}$, whereas $\kappa_\mcm$ 
falls as $r^{-2}$ or more slowly, so $\kappa_\mcm/(-\Delta\kappa)$ 
increases outwards, and the minimum is attained inside the 
image configuration or close to the ellipse boundary.

Second, the observed image multiplicity of every knot must be preserved,
i.e., $N_{\rm im}(\mcm{+}\lambda \mcp)=N_{\rm im}(\mcm)$, which we 
enforce by tracing caustics and bisecting in $\lambda$.

Third, we ensure that the parity of every image is also preserved,
i.e., $P_{\rm im}(\mcm{+}\lambda \mcp)=P_{\rm im}(\mcm)$. This 
condition guarantees that the arrival time order of various images
of a source/knot remains the same.

All of the above conditions require a strong lens model to constrain 
the ShaDe amplitude. If the best-fit model reproduces 
every observed image, the above multiplicity criterion is equivalent
to requiring the observed multiplicity. That said, a best-fit model
can also predict extra images, and the above multiplicity criterion
also considers those images, making it more restrictive than just
considering the observed image position\footnote{On the other hand, 
if the best-fit model ($\mcm$) predicts fewer images than are observed 
for some sources, then the model has failed to produce all observed 
images and we are anchoring the perturbation to the model's own 
failure.}.

\begin{figure*}
    \centering
    \includegraphics[width=0.9\linewidth]{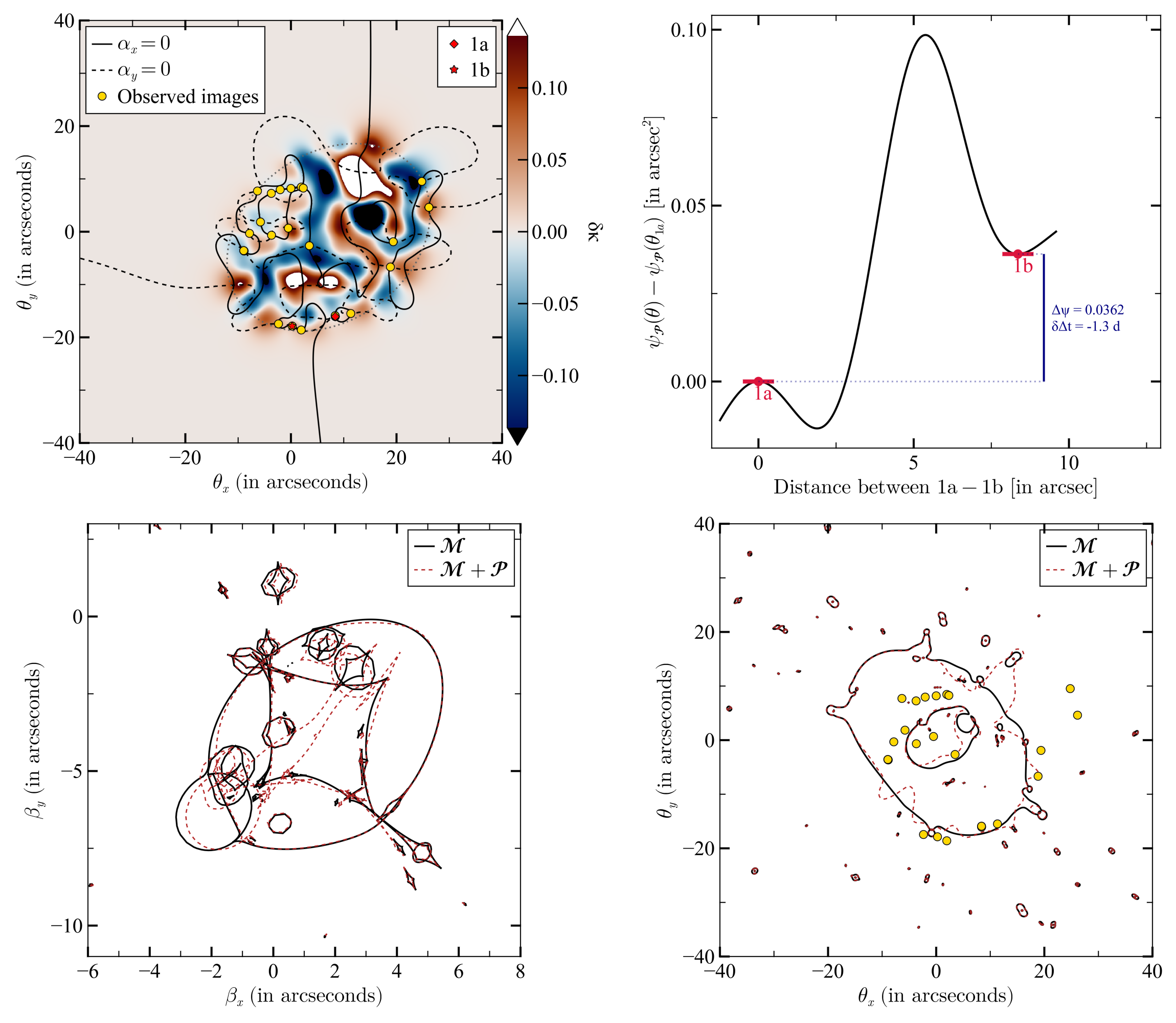}
    \caption{One ShaDes realization for MACS~J0138. 
    \emph{Top left.} Convergence perturbation $\delta\kappa$. Yellow 
    points are the $23$ observed images, and red diamond/star marks 
    the observed images corresponding to SN-Encore. Black solid and 
    dashed curves are the zero contours of $\alpha_{\mcp,x}$ and 
    $\alpha_{\mcp,y}$, which pass through every observed image. 
    \emph{Top right.} The perturbing potential $\psi_\mcp$ along the 
    segment joining 1a and 1b. Its gradient vanishes at both 
    endpoints~(i.e., $\pmb{\alpha}_{\mcp}$) while its value 
    differs between them, shifting $\Delta t_{\rm 1b,1a}$ by 1.3~days 
    through \Eref{eq:key}.
    \emph{Bottom left.} Caustic for the best-fit~($\mcm$) 
    and perturbed~($\mcm+\mcp$) model in solid black and dashed
    red curves, respectively, for~$z_s=1.949$.
    \emph{Bottom right.} Critical curves for the 
    best-fit~($\mcm$) and perturbed~($\mcm+\mcp$) model in solid 
    black and dashed red curves, respectively, for~$z_s=1.949$. The
    yellow points again represent the observed image positions.}
    \label{fig:figure1}
\end{figure*}

\subsection{Effect on time delays}
\label{ssec:shades_td}
For a source at~$z_s$, the arrival time is,
\begin{equation}
    t_d(\pmb{\theta};\pmb{\beta}) = \frac{1+z_d}{\rm c} \frac{D_d D_s}{D_{ds}} 
    \left[\frac{|\pmb{\theta}-\pmb{\beta}|^2}{2} - \frac{D_{ds}}{D_s} \psi_s(\pmb{\theta})\right],
    \label{eq:td}
\end{equation}
where $z_d$ is the lens redshift. $\pmb{\theta}$ and $\pmb{\beta}$
represent the angular positions of image and source, respectively.
A ShaDe leaves $\pmb{\theta}$ and $\pmb{\beta}$ unchanged, so the 
geometric time delay remains invariant, and only the potential time 
delay varies, leading to a net change in the total time delay of,
\begin{equation}
    \delta \Delta t_{d, ij} = - \frac{1+z_d}{\rm c} D_d 
    \left[\psi_\mcp(\pmb{\theta}_j) - \psi_\mcp(\pmb{\theta}_i)\right].
    \label{eq:key}
\end{equation}
We note that the extra delay is independent of source redshifts.
Hence, having multiple source planes does not help as in MSD, 
since \Eref{eq:defining} holds at fiducial normalization for every 
source plane at once. Extra source planes help by contributing extra 
rows to $\mathsf{A}$ and providing constraints at additional points 
in the lens plane. For a Plummer lens, \Eref{eq:key} corresponds to,
\begin{equation}
    \delta\Delta t_{ij}= - \frac{(1+z_{\rm d})}{\rm c} \frac{2\rm G}{\rm c^2}
    \sum_{k} m_k\, \ln\left[\frac{d_{j,k}^2+\theta_{s,k}^2}{d_{i,k}^2 + \theta_{s,k}^2}\right],
    \label{eq:log}
\end{equation}
where $d_{i,k}=|\pmb{\theta}_i-\pmb{\theta}_{c,k}|$. 
\Eref{eq:null} forces the \emph{vector} sum
\begin{equation}
    \pmb{\alpha}_\mcp(\pmb{\theta}_i) = 
    \sum_k m_k\frac{(\pmb{\theta}_i-\pmb{\theta}_{c,k})}{(d_{i,k}^2+\theta_{s,k}^2)},
\end{equation}
to vanish at every image position. However, it says nothing about
the \emph{scalar} sum in \Eref{eq:log}.

The effect of ShaDes on time delays is expected to depend on the 
separation ($L$) of image pairs. For a close pair, the effect is 
suppressed by stationarity. Since 
$\nabla\psi_\mcp = \pmb{\alpha}_\mcp = 0$ at every constrained 
image, $\psi_\mcp$ is stationary at both ends of the segment joining
any two of them. Writing $f(u)=\psi_\mcp$ along that segment,
$u\in[0,L]$, the condition $f'(0)=0$ removes the constant term, 
while $f'(L)=0$ does not remove the curvature but fixes it, 
$f''(0)=-\tfrac12 f'''(0)L$. The two surviving leading order 
contributions to the below
integral, $-\tfrac14 f'''(0)L^3$ and $+\tfrac16 f'''(0)L^3$, 
partially cancel, leaving
\begin{equation}
\begin{aligned}
\psi_\mcp(\pmb{\theta}_j)-\psi_\mcp(\pmb{\theta}_i) & = \int_0^L f'\,{\rm d}u \\
 & = -\tfrac{1}{12}\,f'''(0)\,L^{3} \;+\; \mathcal{O}(L^{4}).
\end{aligned}
\label{eq:cubic}
\end{equation}

Image density further suppresses the effect of ShaDes at small 
separations. If a close pair lies inside a small group of lensed 
images, $\psi_\mcp$ must be stationary at every member of the 
group simultaneously, driving it towards a constant across the 
whole group, 
i.e., $\Delta\kappa = \tfrac12\nabla^2\psi_\mcp \simeq 0$
\cite{Williams2026}. 
Hence, delays measured within such a group would be protected from
ShaDes. However, the same will not be true for two widely 
separated images lying in two different groups, since the 
corresponding potential would be
$\psi_\mcp \simeq C_1$ and $\psi_\mcp \simeq C_2$, and $C_1$ and 
$C_2$ are uncorrelated. By \Eref{eq:key}, the delay between them 
shifts by $-[(1+z_{\rm d})D_{\rm d}/c](C_2-C_1)$, which need not 
be small. A cluster can therefore be image-rich and still carry
a considerable effect on the long-baseline delay from ShaDes. 
However, because $\psi_\mcp$ is itself bounded by the positivity and 
multiplicity conditions~(see~\Sref{ssec:shades_bound}), the ShaDe 
amplitude cannot be arbitrarily large. On the other hand, the 
best-fit/observed time delay would continue to grow with separation,
ultimately decreasing the fractional errors.

\begin{figure}
    \centering
    \includegraphics[width=0.9\linewidth]{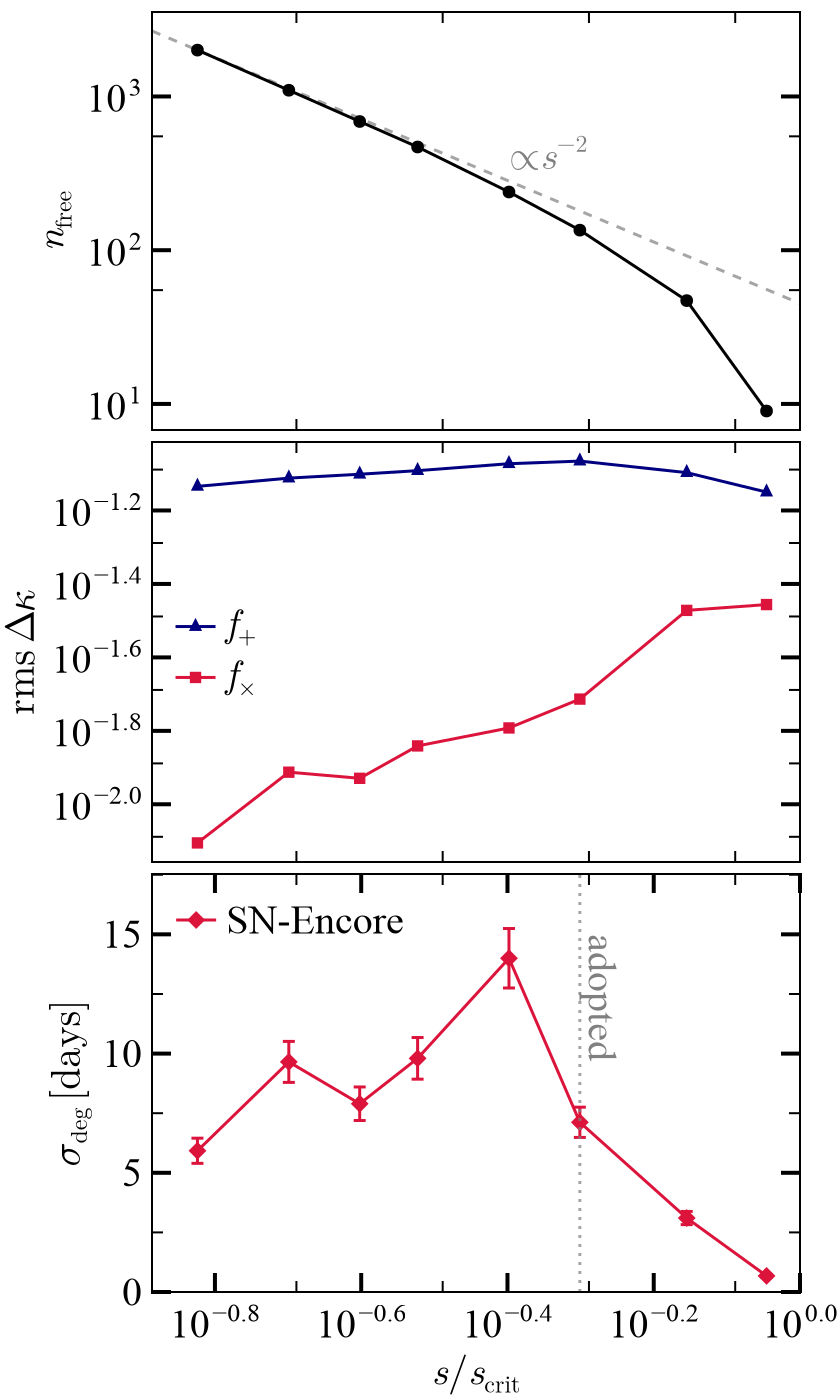}
    \caption{ShaDe dependence on the resolution of the basis.
    \emph{Top panel}. The dimension of the degeneracy space grows as 
    $n_{\rm free}\propto s^{-2}$, Eq.~\eqref{eq:dim}, until the rank 
    of $\mathsf{A}$ saturates at $2N$. 
    \emph{Middle panel}. The perturbation amplitude in rms 
    $\Delta\kappa$, after the positivity cap~($f_+$) alone~(blue curve) 
    and after the multiplicity cap~($f_\times$) as well~(red curve). 
    Positivity is nearly independent of resolution, whereas 
    multiplicity is not, and the widening gap between the curves is 
    the amplitude it removes.
    \emph{Bottom panel}. The resulting $\sigma_{\rm deg}$ for the 
    detected SN-Encore pair, with sampling errors for $64$~realizations
    per point. The dotted line marks the basis scale adopted in
    the current work.}
    \label{fig:figure2}
\end{figure}

\section{MACS\,J0138.0$-$2155}
\label{sec:macs0138}
MACS~J0138.0$-$2155~(MACS~J0138 hereafter; $z_d=0.336$) is the only 
cluster known to lens two supernovae from a common host at $z_s=1.949$, 
SN-Requiem~\cite{2021NatAs...5.1118R} and 
SN-Encore~\cite{2025A&A...702A.157E, 2025A&A...697A..94G, 2026ApJ...998..219P}, 
with SN-Requiem expected to reappear within a year from 
now~\cite{2026A&A...708A.291S}.
In this section, we construct the ShaDes for MACS~J0138 and inspect 
their effects on the time delays. We start by constructing a 
parametric strong lens model with 
\texttt{LensFactory.jl}\footnote{Following, \citet{2026A&A...708A.291S},
we also use flat $\Lambda$CDM cosmology with 
parameters,~$H_0=70\,{\rm km\,s^{-1}\,Mpc^{-1}}$, $\Omega_{m,0}=0.3$ 
and~$\Omega_{\Lambda}=0.7$.}~\cite{2026arXiv260908649M}, 
as we need it to compute
bounds on ShaDe amplitude. Our best-fit lens model has an rms 
of~$0.25''$. Since the underlying full data set is yet to be made 
publicly available, we restrict ourselves from providing any additional 
details on the data products\footnote{In future, the lens model will
be made available at: 
\url{https://github.com/akmeena766/ClusterLensModels}}.
Since convergence is one of the quantities that is 
constrained well in lens modeling~\cite{2017MNRAS.472.3177M}, using 
a different (parametric) lens model is not expected to change our 
results considerably.
We use the lensed image catalog presented 
in~\citet{2026A&A...708A.291S} containing 23 images corresponding to
6 sources. We use the observed image position to construct ShaDe
realizations.

In total, we construct 500 realizations of ShaDes all satisfying the
first two amplitude bounds. Out of these, we choose 214 realizations 
for our analysis, where parities are also preserved for each image. 
For a given realization, for any image pair of a source, 
$\mathcal{E}=\Delta t^{\mcr}-\Delta t^{\mcm}$, 
where~$\mcr=\mcm+\mcp$, represents the extra time delay due to ShaDes.
The error corresponding to full ensemble of 214 realizations would 
be~$\sigma_{\rm deg} = {\rm std_\mcr(\mathcal{E})}$. 
We note that $\sigma_{\rm deg}$ is a property of the ShaDes and 
underlying image configuration. 
Since $H_0 \propto \Delta\phi_{ij}/\Delta t^{\rm obs}_{ij}$ with
$\Delta\phi_{ij}$ representing the Fermat potential, at 
fixed measured delay
\begin{equation}
\frac{\delta H_0}{H_0} = f_{\rm deg} \coloneq \frac{\sigma_{\rm deg}}{|\Delta t|_{\rm obs}}.
\label{eq:H0_error}
\end{equation}
Here, we have used observed time delays. However, it can be replaced
with model predicted delays corresponding to any model.

For SN-Encore, the measured/observed time delay 
is~$\Delta t_{\rm 1b,1a} = -39.8^{+3.9}_{-3.3}$~days~\cite{2026ApJ...998..219P},
a short, close-separation pair.
Our~$\sigma_{\rm deg}$ for this pair is $\simeq7$~days, i.e., $17.5\%$
of the measured time delay, which by \Eref{eq:H0_error} is equivalent
to $\simeq17.5\%$ systematic on the inferred ~$H_0$ value.
\citet{2026A&A...708A.291S} inferred~$H_0=66.9_{-8.1}^{+11.2}\,{\rm km/s/Mpc}$
with~$+17\%$ and~$-12\%$ errors dominated by the uncertainty in
measured time delay. The ShaDe error is not incorporated in the 
above error. The initial ShaDe amplitudes are drawn randomly, but
the amplitudes are fixed by the condition discussed in 
\Sref{ssec:shades_bound}. Hence, it would not be correct to assume
that all distributions are independent and random. However, at the
same time, the ShaDes can be added to any of the seven lens models
presented in \citet{2026A&A...708A.291S}. In that case, we can add
the corresponding errors in quadrature.
\Fref{fig:figure1} shows one ShaDe realization for
MACS~J0138 and the corresponding change in the potential time delay 
for SN-Encore. In the image plane (top left panel), we note
that the presence of a higher density of lensed images suppresses
the ShaDe amplitude in the upper left part. However, that is not
the case for the host arc of SN-Encore and SN-Requiem.

From all the observed lensed images taken together, we construct 
a single matrix $\mathsf{A}$ (Eq.~\ref{eq:null}), and hence a 
single $\ker[\mathsf{A}]$ for 
the whole cluster. SN-Requiem, which lies in the same host, is 
also subjected to the same degeneracy space.
\citet{2026A&A...708A.291S} predict the next images of SN-Encore
and SN-Requiem to appear with delays~$\gtrsim3000$~days and
$\sim 3700-4000$~days, respectively, leading to an error of $2\%-3\%$
on inferred $H_0$ value. For this long-baseline pair of 
five-imaged host, we find $\sigma_{\rm deg}\simeq60$~days leading 
to the ShaDe contribution of~$1.5\%-2.0\%$ on $3,000-4,000$~days 
intervals. 
Again, it is important to note that the time delay for the next 
image of SN-Requiem is expected to be measured at the percent 
level, and the $2\%-3\%$ error corresponds to the modeling and 
does not include the ShaDe contribution. Inferences for SN-Requiem 
are based on host images rather than on its model-predicted images, 
so as not to interfere with the ongoing analysis for this cluster. 
However, we can expect similar errors for the model-predicted image, 
as we quote numbers for a range of time delays.

\section{Dependence on the basis}
\label{sec:shades_basis}
From \Eref{eq:dim}, we note that the degeneracy dimensions go
as~$n_{\rm free}\propto s^{-2}$.
Hence, one can argue that~$\sigma_{\rm deg}$ is set by resolution 
of the basis rather than the data itself. In \Fref{fig:figure2}, we 
vary~$s/s_{\rm crit}$ over a factor of six and the~$n_{\rm free}$ 
varies by more than two orders of magnitude, as shown in the top 
panel. However, the admissible degeneracy amplitude falls, as we can 
see in the middle panel. For $s \leq 0.5 s_{\rm crit}$, 
$\sigma_{\rm deg}$ varies only by
a factor of two, without a systematic trend towards finer basis.
We also note that the adopted~$s=s_{\rm crit}/2$ is not the 
worst-case in terms of uncertainties for SN-Encore. 
At~$s=0.4\,s_{\rm crit}$, $\sigma_{\rm deg}$ increases to 
$\sim14$~days, i.e., a factor two above the quoted values.

The middle panel in \Fref{fig:figure2} also highlights the 
importance of the multiplicity cap over the positivity cap. 
The positivity cap is nearly blind to the basis scale holding 
$\rm{rms}\,\Delta\kappa\simeq0.08$. On the other hand, the 
multiplicity cap steadily decreases the $\rm{rms}\,\Delta\kappa$ as
we increase the number of basis components. This implies that a 
finer basis is efficient in creating/removing extra images, which 
is what the multiplicity cap sets. Hence, it is the multiplicity 
that actually bounds the perturbations rather than the positivity 
of convergence.

Below $s/s_{\rm crit} = 0.25$, we begin to encounter directions 
in $\ker[\mathsf{A}]$ for which no amplitude preserves the 
observed multiplicities. Such directions are excluded by the
data entirely, rather than merely bounded, and their number grows 
as the basis is refined. The multiplicity condition, therefore, 
removes not only amplitude but dimension, so that $n_{\rm free}$
increasingly overstates the freedom the data actually permit.

\section{Constant deflection}
\label{ssec:constant}
In the above sections, by definition, the deflection at the 
observed image positions is zero. However, we note that to preserve 
the observed image positions, zero deflection is not a necessary 
condition. We can have a constant deflection at image positions and 
shift the corresponding source positions such that \Eref{eq:defining}
becomes,
\begin{equation}
    \pmb{\alpha}_\mcp(\pmb{\theta}_i) = \mathbf{c}_s, \quad \forall\, s, \;\; \forall\, i \in s,
    \label{eq:shades_gen}
\end{equation}
where $i$ represents images and $s$ represents the sources. Hence,
$\pmb{c}_s$ is a vector of length equal to the number of lensed 
images and has the same values for images belonging to the same source. 
The corresponding shift in the source position 
is~$\pmb{\beta}_s \to \pmb{\beta}_s - (D_{ds}/D_s) \pmb{c}_{s}$. 
This is a more general version of \Eref{eq:defining} and increases the 
number of dimensions in the degeneracy space. This raises the 
question: How does this affect the time delay uncertainty estimated
above? We find that, at least, for the observed pair of SN-Encore, 
even after including the constant source offset, the uncertainty in
corresponding time delay remains~$\simeq7$~days. This can be 
attributed to the fact that the host of SN-Encore lies close to the
caustic in the source plane and does not have a large freedom in its 
position while preserving the image multiplicity for every knot. For
the upcoming image of SN-Requiem, the uncertainty in time delay grows
to~$\simeq112$~days, which translates to~$2.8\%-3.7\%$ error in $H_0$
assuming a time delay of $3,000-4,000$~days.

\section{Discussion}
What makes ShaDes notorious is not their size but their durability.
Extra source planes break the MSD because it rescales $\pmb{\beta}$ 
by a factor that cannot be correct for all source redshifts 
simultaneously. 
In ShaDes, \Eref{eq:defining}, holds at fiducial normalization and is 
satisfied for every plane at once, so additional sources help only 
through the extra lensed image positions they contribute, as extra 
constrain rows of $\mathsf{A}$. Comparing independent lens models does 
not help.
The spread between models measures the variance induced by choices 
that differ between teams, but every team constrains its model
with image positions, so every team sits somewhere in the affine 
space $\{\mcm + \mcp : \mcp\in\ker[\mathsf{A}]\}$ with no data-driven 
means of choosing where. Their mutual scatter is the projection of 
their disagreement onto the constrained directions, the complement of
the space explored here. Even image density, which does suppress the 
degeneracy, does so only locally and only for pairs inside one tight 
group.

For SN-Encore, we find that ShaDes introduce an extra time delay
of $\delta t_d \simeq 7$~days on $\Delta t_{\rm 1b,1a}$, which is
$17.5\%$ of the measured time delay. By
\Eref{eq:H0_error} this is a $17.5\%$ systematic on $H_0$.
SN-Requiem, in the same host and therefore behind the same
$\ker[\mathsf{A}]$, is also expected to be affected in the same way.
For SN-Requiem, the systematic uncertainty on the delays available 
today is of the same order.

The reappearance is the favorable case. The next images of SN-Encore 
and SN-Requiem arrive on the baselines of 
$3,000$--$4,000$~days, and ShaDes contributes only 
$60$~days (i.e., $1.5$--$2.0\%$) and the error increase 
to~$\simeq112$~days (i.e., $2.8$--$3.7\%$) once constant deflection 
at the image positions is introduced. 
That is not negligible against the $2$--$3\%$ $H_0$ precision 
those measurements are intended to reach~\cite{2026A&A...708A.291S},
but it is an order of magnitude smaller, fractionally, than the 
effect on the delays already in hand.
While the current system being measured today seems to exhibit a 
severe ShaDes systematic error, the recurrence of SN-Requiem will 
not.

Our numbers about ShaDes should be treated as floor values, and 
it is worth being precise about why. 
Two choices were made conservatively, and none of them is 
expected to lower $\sigma_{\rm deg}$. 
First, we imposed the strict condition \Eref{eq:defining}, so our 
perturbations do not move the images. A lens model that fits image 
positions within a positional uncertainity $\sigma_{\rm pos}$ 
admits the entire sublevel 
set $\|\mathsf{A}\mathbf{m}\|\lesssim\sqrt{2N}\,\sigma_{\rm pos}$, 
and $\ker[\mathsf{A}]$ is a strict subset of it for every
$\sigma_{\rm pos}>0$. 
Second, the amplitude of each realization is set by the single most
restrictive pixel in \Eref{eq:pos}, so one location caps the whole
map. Relaxing any of the above criteria would enlarge the admissible 
family, so the true systematic is larger than what we quote. 
The effect of these choices on~$H_0$ measurements will be presented
in a forthcoming paper. 

Placing a lensed image in a previously unconstrained part of the 
field removes two dimensions from the null space. Hence, pixel-level 
modeling ~\citep[e.g.,][]{2024ApJ...976..110A, 2025MNRAS.544..708X, 
2026ApJ..1003..146E} of an extended lensed arc can significantly 
decrease the ShaDe amplitude, at least in the surrounding regions.
However, this is not a generic fix, and one needs to dig deeper~(on
a case-by-case basis) to check the extent to which ShaDes survive 
in extended source modeling. In addition, the lensed arcs imply
lensed images close to the critical curve, regions where various 
lens models can have much larger systematic 
errors~\cite{2017MNRAS.472.3177M}, again highlighting the requirement
for a more detailed analysis of ShaDes in extended source modeling
with multiple lens models.

One obvious objection to ShaDes is that they may not correspond to
physical features in the lens, as they are signed components laid on
a lattice with no reference to the cluster light. A modeler working 
under the light trace mass~(LTM) assumption would never produce a 
ShaDe, and might argue that their prior excludes it. LTM is itself a
modeling assumption, not a measurement, and free-form reconstructions 
routinely recover features that light does not 
trace~\cite{2023MNRAS.525.2519G, 2024MNRAS.527.2639P, 
2025MNRAS.536.2690P, 2025A&A...703A..10L} and these might be due to 
ShaDes~\cite{Williams2026}.
Settling how much of $\ker[\mathsf{A}]$ survives an explicit LTM prior 
requires re-drawing $\mathbf{m}$ keyed to the cluster light and 
reporting $\sigma_{\rm deg}$ as a function of prior strength, 
which we defer to future work.

\section*{Acknowledgments}
AKM acknowledges support from the Start-up Grant IE/CARE-25-0305 
provided by the IISc, Bengaluru, India.
This research has made use of NASA's Astrophysics Data System 
Bibliographic Services. AKM acknowledges the use of Claude~Opus~5 
for code and grammar assistance.

\section*{Data Availability}
The software packages used in this work, 
\href{https://github.com/akmeena766/LensFactory.jl}{\texttt{LensFactory.jl}}~\cite{2026arXiv260908649M} and 
\href{https://github.com/akmeena766/ShaDes.jl}{\texttt{ShaDes.jl}}\footnote{\url{https://github.com/akmeena766/ShaDes.jl}}
are publicly available. The data generated in this study can be made
available on a reasonable request to the corresponding author.






\def\mnras{MNRAS}
\def\aap{A\&A}
\def\aj{AJ}
\def\apj{ApJ}
\def\apjl{ApJL}
\def\apjs{ApJS}
\def\nar{Nature Ast}
\def\physrep{Phys. Rep.}
\def\ssr{Space Sci. Rev.}
\textit{}
\bibliographystyle{apsrev4-2}
\bibliography{Reference} 

\end{document}